\documentclass[letterpaper, 10 pt, conference]{ieeeconf}  

\IEEEoverridecommandlockouts                              

\usepackage{tikz}
\usepackage{amsmath}
\usepackage{mathtools}
\usepackage{bbm}
\usepackage{algorithm,algorithmic}
\usepackage{hyperref}
\usepackage{epsfig}
\usepackage{amssymb}
\usepackage[T1]{fontenc}
\usepackage[utf8]{inputenc}

\title{\LARGE \bf
MAMPS: Safe Multi-Agent Reinforcement Learning via Model Predictive Shielding}

\author{Wenbo Zhang$^{1}$, Osbert Bastani$^{2}$, and Vijay Kumar$^{1}$
\thanks{$^{1}$W. Zhang and V. Kumar are with the General Robotics, Automation, Sensing $\&$ Perception (GRASP) Laboratory, University of Pennsylvania, USA
        }\thanks{$^{2}$O. Bastani is with the Department of Computer and Information Science, University of Pennsylvania, USA
{\tt\small $\{$zwenbo, obastani, kumar$\}$@seas.upenn.edu}
       } 
}

\begin{document}

\maketitle
\thispagestyle{empty}
\pagestyle{empty}

\begin{abstract}
Reinforcement learning is a promising approach to learning control policies for performing complex multi-agent robotics tasks. However, a policy learned in simulation often fails to guarantee even simple safety properties such as obstacle avoidance. To ensure safety, we propose multi-agent model predictive shielding (MAMPS), an algorithm that provably guarantees safety for an arbitrary learned policy. In particular, it operates by using the learned policy as often as possible, but instead uses a backup policy in cases where it cannot guarantee the safety of the learned policy. Using a multi-agent simulation environment, we show how MAMPS can achieve good performance while ensuring safety.
\end{abstract}

\section{Introduction}

Reinforcement learning~\cite{gu2017deep,andrychowicz2018learning} has been shown to be a promising technique for learning control policies for complex robotics tasks ranging from autonomous vehicles~\cite{tanner2003coordination} to home service robots~\cite{srinivasa2010herb}, or to ``compress'' an expensive model predictive controller (MPC) into a much faster neural network policy~\cite{levine2013guided}. A major challenge in using reinforcement learning is safety~\cite{gillula2012guaranteed,akametalu2014reachability,saha2014automated,fisac2019bridging,liu2017planning,schwarting2017parallel,alshiekh2018safe,aisafety}---control policies learned using reinforcement learning typically do not provide any safety guarantees, even when the safety property is explicitly considered by the learning algorithm.

As a consequence, there has been much interested in algorithms that provide safety guarantees for a learned control policy $\hat{\pi}$. We are interested in the setting where $\hat{\pi}$ is learned in simulation, and we want to ensure safety after it is deployed on a robot (assuming that our model of the dynamics is correct). One approach is to formally prove that $\hat{\pi}$ is safe~\cite{berkenkamp2017safe,bastani2018verifiable,ivanov2019verisig}. An alternative approach, called \emph{shielding}, is to synthesize a backup controller $\pi_{\text{backup}}$ that is guaranteed to be safe on some subset of states, which we call \emph{recoverable states}~\cite{gillula2012guaranteed,akametalu2014reachability,alshiekh2018safe,DBLP:journals/corr/abs-1803-08552,bastani2019safe}. Then, the \emph{shield policy} $\pi_{\text{shield}}$ uses $\pi_{\text{backup}}$ whenever $\hat{\pi}$ would bring the robot to an irrecoverable state on the next step; otherwise it uses $\hat{\pi}$. A key challenge with existing approaches is that they rely on verifying either $\hat{\pi}$ or $\pi_{\text{backup}}$, which typically does not scale to high-dimensional systems. For systems with obstacles in the environment or other agents that must be encoded in the state, verifying safety quickly becomes intractable.

\begin{figure}[t]
\centering
\includegraphics[keepaspectratio, width=0.47\textwidth]{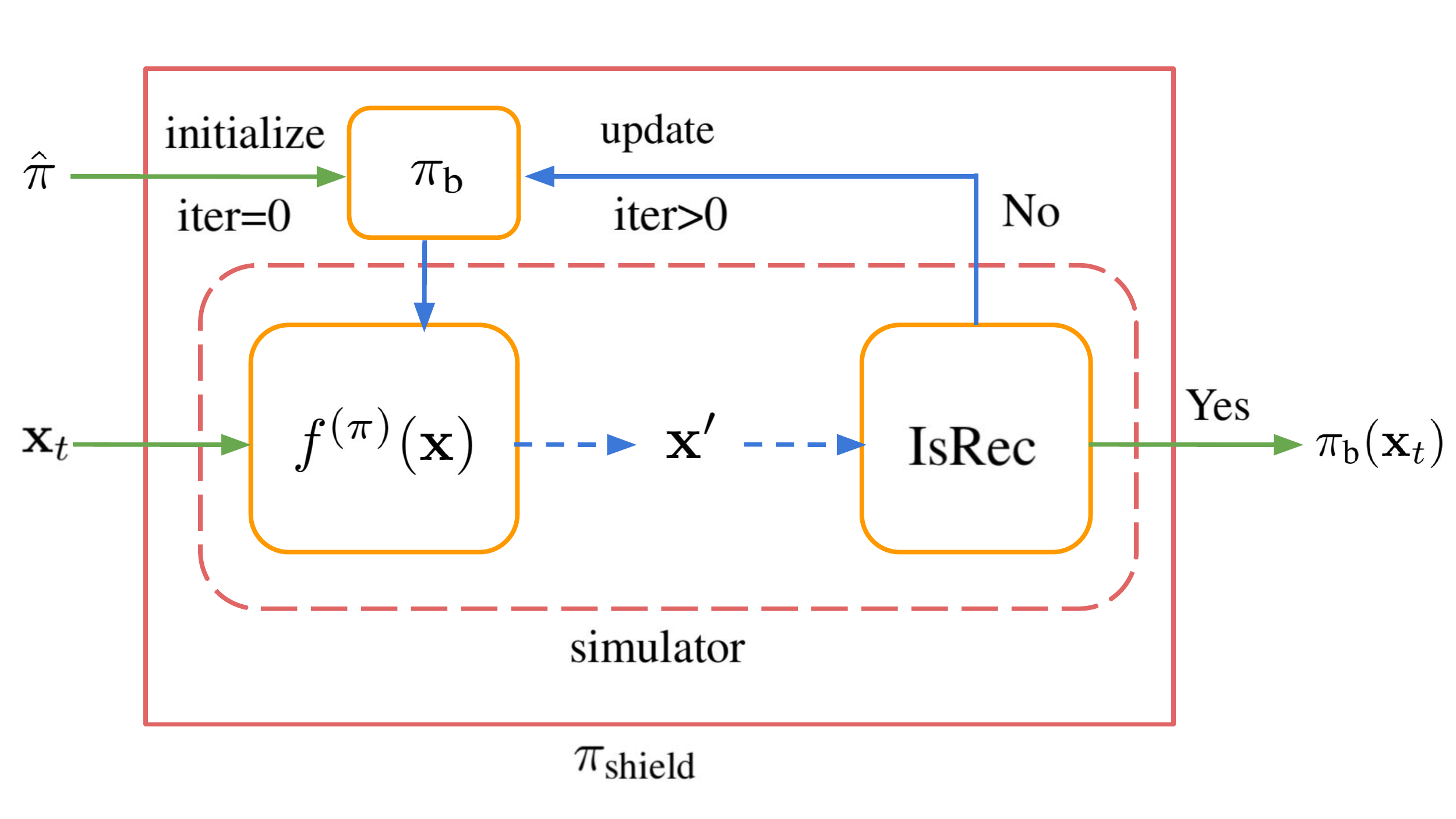}
\caption{\textbf{Overview of multi-agent model predictive shielding (MAMPS).} On step $t$, the current state of the multi-agent system is $\bold{x}_t$. The red solid box shows the entire MAMPS control policy is $\pi_\text{shield}$ module. There are three basic components in MAMPS: the ``current policy'' $\pi_b$, the closed-loop dynamics $f^{(\pi)}(\bold{x})$, and the subroutine IsRecoverable (shorted to IsRec). The current policy is a combination of learned policies and recovery policies for different agents. First, MAMPS initializes $\pi_b$ to $\hat{\pi}$ for each agent. Then, it iteratively determines whether using $\pi_b$ transitions the system to a recoverable state $\bold{x}'$. If not, then it switches agents that were unsafe to $\pi_{\text{rec}}$ (the blue line). Once it has found $b$ such that $f^{(\pi_b)}(\bold{x})$ is recoverable, it can safely output action $\pi_\text{b}(\bold{x}_t)$ (the green line). The red dashed box shows the internal simulation used by MAMPS to check recoverability; note that all dynamics applications in this box are according to a simulation run by the MAMPS algorithm, not according to the true dynamics. The true dynamics are only applied once $\pi_{\text{shield}}$ returns an action $\pi_b(\bold{x}_t)$.}
\label{fig:workflow}
\end{figure}

A promising alternative is \emph{model predictive shielding} (MPS)~\cite{DBLP:journals/corr/abs-1803-08552,bastani2019safe,li2019robust}, which performs shielding on-the-fly instead of ahead-of-time. The intuition is that checking whether a single state is recoverable (i.e., the next one) is much more efficient, even on-the-fly, than exhaustively checking recoverability for all states ahead-of-time. In particular, to check whether a given state is recoverable, MPS simply simulates the closed-loop dynamics (assumed to be deterministic) with $\pi_{\text{backup}}$ and checks whether it is safe. While we focus on deterministic dynamics, there has been recent work extending MPS to stochastic dynamics~\cite{li2019robust}.

In this paper, we study the safety problem for multi-agent systems~\cite{desai2001modeling,alonso2015multi}. In this setting, approaches to safety have been proposed based on restricting the velocities of the agents~\cite{AlonsoMora2010OptimalRC,rvo,khan2019learning}; however, these approaches only apply to systems that are holonomic~\cite{rvo,khan2019learning} or approximately holonomic~\cite{AlonsoMora2010OptimalRC}. Instead, we propose an approach based on MPS, which applies to general robot dynamics. We can in principle treat the multi-agent system as a high-dimensional single-agent system, and then apply MPS to ensure safety. However, this approach can achieve very suboptimal results since even if a single robot is about to be irrecoverable, then the shield uses $\pi_{\text{backup}}$ for every single robot in the system.

We propose an approach, called multi-agent model predictive shielding (MAMPS), summarized in Fig.~\ref{fig:workflow}, which avoids this problem by incrementally switching each agent from using $\hat{\pi}$ to using $\pi_{\text{backup}}$. The iterative process is needed since switching one agent to using $\pi_{\text{backup}}$ can cause other agents to subsequently become unsafe. The general workflow is shown in Fig. \ref{fig:workflow}. We prove that this modification preserves the key safety guarantees provided by shielding---i.e., if the system starts in a recoverable state, then using $\pi_{\text{shield}}$ keeps the system safe over an infinite horizon.

\textbf{Contributions.}
Our key contributions are: (i) the multi-agent model predictive shielding (MAMPS) algorithm for ensuring safety of a learned policy (Section~\ref{sec:formulation} \&~\ref{sec:MPS_IN_MA}), (ii) a theoretical guarantee regarding the safety of this algorithm (Section~\ref{sec:MPS_IN_MA}), and (iii) an experimental evaluation demonstrating how this algorithm outperforms the standard MPS algorithm in the multi-agent setting (Section~\ref{sec:experiments}).

\section{Problem Formulation}
\label{sec:formulation}

In this section, we formulate the problem of shielding a learned policy to ensure safety.

\textbf{Dynamical system.}
We consider a deterministic, discrete time dynamical system with continuous states $\bold{x} \in \mathcal{X} \subseteq \mathbb{R}^{n_X}$, continuous action $\bold{u}^i \in \mathcal{U}^i\subseteq\mathbb{R}^{n_U}$ (for each agent $i \in [N] = \{1,2,...,N\}$), dynamics $f: \mathcal{X} \times \mathcal{U}^1 \times \dots \times \mathcal{U}^N \to \mathcal{X}$, and probability distribution $d_0$ over initial states. We assume that $\bold{x}\in\mathcal{X}$ encodes the state of all the agents as well as goals and obstacles in the environment.

As a running example, consider a system of non-holonomic robots with acceleration control and steering control---i.e., each robot $i\in[N]$ has state $\bold{x}^i=(x^i, y^i, v^i, \theta^i)$ and action $\bold{u}^i=(a^i, \delta^i)$, where $(x^i,y^i)$ is the position of the robot in $\mathbb{R}^2$, $v^i$ is its velocity, $\theta^i$ is its heading, $a^i$ is its acceleration, and $\delta^i$ is its steering angle. In addition, we consider $N$ goals with positions $\bold{g}^i=(g^i,h^i)\in\mathbb{R}^2$ for each $i\in[N]$, as well as $M$ obstacles with positions $\bold{z}^i=(z^i,w^i)\in\mathbb{R}^2$ for each $i\in[M]$. The goal is for agent $i$ to reach goal $i$. Together, the multi-agent state is $\bold{x}=(\bold{x}^1, ..., \bold{x}^N,\bold{g}^1,...,\bold{g}^N,\bold{z}^1,...,\bold{z}^N$).

\textbf{Control policy.}
Given a vector of control policies $\pi=(\pi^1,...,\pi^N)$, where $\pi^i:\mathcal{X}\to\mathcal{U}^i$ for each $i\in[N]$, we use $f^{(\pi)}(x,u)=f(x,\pi^1(x),...,\pi^N(x))$ to denote the closed-loop dynamics. An infinite-horizon trajectory generated using $\pi$ from initial state $\bold{x}_0\in\mathcal{X}$ is the sequence $\zeta=(\bold{x}_0,\bold{x}_1,...)$, where $\bold{x}_{t+1}=f^{(\pi)}(\bold{x}_t)$. Similarly, given a finite horizon $T\in\mathbb{N}$, a finite-horizon trajectory generated using $\pi$ from initial state $\bold{x}_0\in\mathcal{X}$ is the sequence $\zeta=(\bold{x}_0,\bold{x}_1,...,\bold{x}_{T-1})$.

\textbf{Safe states.}
We assume given sets $\mathcal{X}_{\text{safe}}\subseteq\mathcal{X}$ indicating that agent $i\in[N]$ is safe. Then, the system as a whole is safe if every agent is safe---i.e.,
\begin{align*}
\mathcal{X}_{\text{safe}}=\bigcap_{i=1}^N\mathcal{X}_{\text{safe}}^i.
\end{align*}
The goal is to ensure that the system never transitions to an unsafe state $\bold{x}\not\in\mathcal{X}_{\text{safe}}$. Given a trajectory $\zeta=(\bold{x}_0,\bold{x}_1,...)$ we say $\zeta$ is safe if $\bold{x}_t\in\mathcal{X}_{\text{safe}}$ for all $t\ge0$. 

In our example, an agent $i\in[N]$ is safe if it has not collided with an obstacle or any other robots. In particular, we have $\bold{x}\in\mathcal{X}_{\text{safe}}^i$if it satisfies the following constraints:
\begin{align*}
d(\bold{x}^i, \bold{x}^j) & \geq 2r_\text{robot} + m \hspace{0.1in} (\forall j \in [N],~i\neq j) \\
d(\bold{x}^i, \bold{z}^j) & \geq r_\text{robot} + r_\text{obstacle} + m \hspace{0.1in} (\forall j \in [M]),
\end{align*}
where $r_{\text{robot}}$ is the robot radius, $r_{\text{obstacle}}$ is the obstacle radius, $m$ is a safety margin, and $d$ is the Euclidean distance. As a consequence, the overall system is safe if the above constraints are satisifed for all $i\in[N]$.

\textbf{Stable controller and stable states.}
To ensure safety for an infinite horizon, we assume that we are given a control policy $\pi_{\text{stable}}$ and a subset $\mathcal{X}_{\text{stable}}\subseteq\mathcal{X}_{\text{safe}}$ of states such that using $\pi_{\text{stable}}$ guarantees safety indefinitely. In particular, we assume given sets $\mathcal{X}_{\text{stable}}^i\subseteq\mathcal{X}_{\text{safe}}^i$ indicating that agent $i\in[N]$ is stable. Then, the system as a whole is stable if every agent is stable---i.e.,
\begin{align*}
\mathcal{X}_{\text{stable}}=\bigcap_{i=1}^N\mathcal{X}_{\text{stable}}^i.
\end{align*}
Our key assumption is that for any $\bold{x}\in\mathcal{X}_{\text{stable}}$, the trajectory $\zeta$ generated using $\pi_{\text{stable}}$ from initial state $\bold{x}_0=\bold{x}$ is safe.

In our running example, we have $\bold{x}\in\mathcal{X}_{\text{stable}}^i$ if $v^i=0$ and $\bold{x}\in\mathcal{X}_{\text{safe}}$---i.e., agent $i$ is at rest and the overall system is safe. Furthermore, the backup control policy is $\pi^i(x)=(0,0)$---i.e., the steering angle and acceleration are both zero. As a consequence, we have $\bold{x}\in\mathcal{X}_{\text{stable}}$ if $v^i=0$ for every agent $i\in[N]$ and furthermore $\bold{x}\in\mathcal{X}_{\text{safe}}$. In other words, in a stable state, all agents are at rest, and the backup control policy keeps them at rest.

\textbf{Reward.}
We consider a reward function $r:\mathcal{X}\times\mathcal{U}\to\mathbb{R}$ that we seek to maximize. Then, given a finite time horizon $T$, initial states $\mathcal{X}$ and an initial state distribution $d_0$, our goal is to find a policy $\hat{\pi}$ that maximizes
\begin{align*}
\mathbb{E}_{\bold{x}_0\sim d_0}\left[\sum_{t=0}^{T-1}r(\bold{x}_t,\bold{u}_t)\right],
\end{align*}
where $\bold{x}_{t+1}=f^{(\pi)}(\bold{x}_t,\bold{u}_t)$ and $\bold{u}_t=\pi(\bold{x}_t)$. In our example, the reward might be to minimize the distance of each robot to a goal state:
\begin{align*}
r(\bold{x},\bold{u})=-\sum_{i=1}^Nd(\bold{x}^i,\bold{g}^i),
\end{align*}
where $\bold{g}^i\in\mathbb{R}^2$ is a goal state.
\footnote{For our experiments, we use a more complex reward function; see Section \ref{sec:experiments} for details.}
The rewards can also encode soft constraints that encourage the robot to remain in $\mathcal{X}_{\text{safe}}$.

\textbf{Learned policy.}
We assume given a vector of policies $\hat{\pi}=(\hat{\pi}^1,...,\hat{\pi}^N)$, where each policy can be arbitrary---e.g., $\hat{\pi}$ can be learned using the multi-agent deep deterministic policy gradient algorithm (MADDPG)~\cite{lowe2017multi, mordatch2017emergence}.

\textbf{Shielding problem.}
A policy $\pi$ is safe if for any trajectory $\zeta=(\bold{x}_0,\bold{x}_1,...)$ starting from $\bold{x}_0\in\mathcal{X}_{\text{stable}}$, $\zeta$ is safe.
\footnote{We restrict to trajectories starting from stable states; it is impossible in general to guarantee safety for trajectories starting from unstable states without leveraging specific aspects of the dynamical system in question.}
Then, our goal is to construct a vector of control policies $\pi_\text{shield}$ that is safe. To try and maximize reward, our construction of $\pi_{\text{shield}}$ leverages $\hat{\pi}$---in particular, it tries to maximize the number of states $\bold{x}\in\mathcal{X}$ for which $\pi_{\text{shield}}(\bold{x})=\hat{\pi}(\bold{x})$.

\begin{algorithm}[t]
\caption{Compute the MAMPS policy for state $\bold{x}$.}
\label{alg:shield}
\begin{algorithmic}[1]
\renewcommand{\algorithmicrequire}{\textbf{function}}
\REQUIRE 
MAMPS($\bold{x}$):
\STATE $b\gets(\text{true},...,\text{true})$
\WHILE{true}
\STATE $\eta_{\text{rec}}\gets\text{IsRecoverable}(f^{(\pi_b)}(\bold{x}))$
\IF{$\eta_{\text{rec}}=(\text{true},...,\text{true})$}
\RETURN $f^{(\pi_b)}(\bold{x})$
\ENDIF
\FOR{$i\in[N]$}
\STATE $b_i'\gets b_i\wedge\eta_{\text{rec}}^i$
\ENDFOR
\IF{$b'=b$}
\STATE $b\gets(\text{false},..,\text{false})$
\ELSE
\STATE $b\gets b'$
\ENDIF
\ENDWHILE
\renewcommand{\algorithmicensure}{\textbf{Output:}} 
\end{algorithmic} 
\end{algorithm}

\begin{algorithm}[t]
\caption{Check if each agent is recoverable at state $\bold{x}$.}
\label{alg:recovery}
\begin{algorithmic}[1]
\renewcommand{\algorithmicrequire}{\textbf{function}}
\REQUIRE IsRecoverable($\bold{x}$):
\STATE $\eta_{\text{safe}}\gets(\text{true},...,\text{true})$
\STATE $\eta_{\text{rec}}\gets(\text{false},...,\text{false})$
\FOR{$t \in \{0,1,...,T_{\text{max}}-1\}$}
\FOR{$i\in[N]$}
\STATE $\eta_{\text{safe}}^i\gets\eta_{\text{safe}}^i\wedge(\bold{x}\in\mathcal{X}_{\text{safe}}^i)$
\STATE $\eta_{\text{rec}}^i\gets\eta_{\text{rec}}^i\vee\Big(\eta_{\text{safe}}^i\wedge(\bold{x}\in\mathcal{X}_{\text{stable}}^i)\Big)$
\ENDFOR
\STATE  $\bold{x} \gets f^{(\pi_\text{backup})}(\bold{x})$
\ENDFOR
\RETURN $\eta_{\text{rec}}$
\renewcommand{\algorithmicensure}{\textbf{Output:}} 
\end{algorithmic} 
\end{algorithm} 

\begin{figure}[t]
\centering
\includegraphics[keepaspectratio, width=0.45\textwidth]{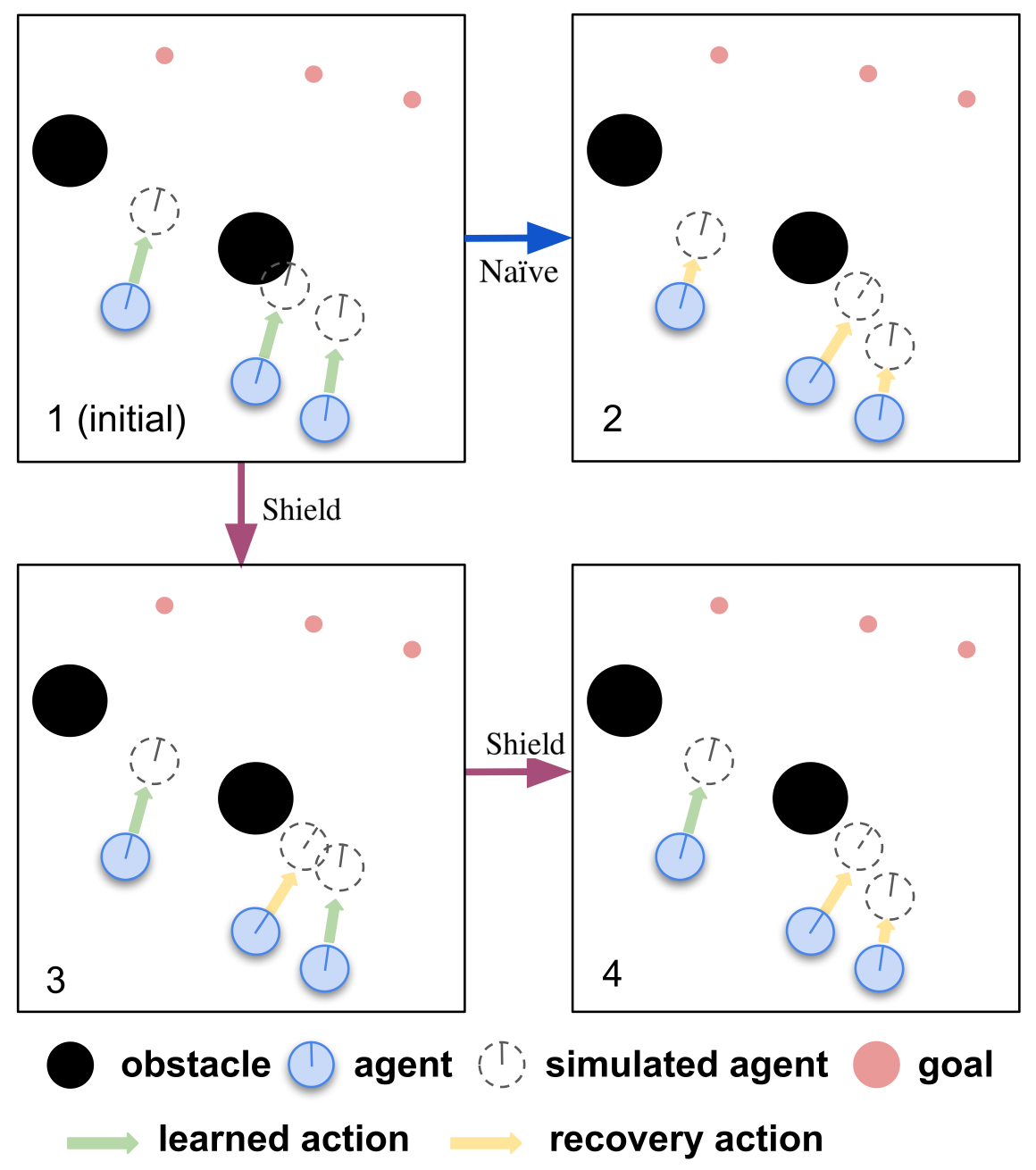}
\caption{\textbf{Comparison of MAMPS to the na\"{i}ve approach.} Image 1 is the initial state. In this state, the middle agent will collide with the obstacle if it uses the learned policy $\hat{\pi}$. The na\"ive approach treats the system as a single-agent system and uses MPS~\cite{bastani2019safe}. In this approach, all agents will switch to using the recovery policy $\pi_{\text{rec}}$ (Image 2). In contrast, when MAMPS iteratively checks whether successively more conservative configurations can transition the system to a recoverable state. In this example, MAMPS first switches the middle agent to using the recovery policy (Image 3). However, this change causes a new problem---the middle agent will now collide with the right-most agent. Thus, in the second iteration, MAMPS switches the right-most agent to using the recovery policy (Image 4). This choice transitions the system to a recoverable state, so MAMPS returns these actions. In this example, MAMPS allows the left-most agent to use the learned policy, whereas the na\"{i}ve approach switches it to the recovery policy. Thus, MAMPS can achieve a significantly higher reward.}
\label{fig:naive_and_ours}
\end{figure}

\section{Multi-Agent Model Predictive Shielding}
\label{sec:MPS_IN_MA}

We propose an extension of model-predictive shielding (MPS)~\cite{bastani2019safe} to the setting of multi-agent systems, which we call multi-agent model predictive shielding (MAMPS).

\textbf{Background on MPS.}
Recall that $\pi_{\text{stable}}$ can guarantee safety for an infinite horizon starting from any stable state $\bold{x}\in\mathcal{X}_{\text{stable}}$. Thus, we can use the shielding approach using $\pi_{\text{stable}}$. In particular, suppose we start at a state $\bold{x}\in\mathcal{X}_{\text{stable}}$. To decide whether to use $\hat{\pi}$ or $\pi_{\text{stable}}$, we check if $\bold{x}'=f^{(\hat{\pi})}(\bold{x})\in\mathcal{X}_{\text{stable}}$. If so, then it is safe to use $\hat{\pi}(\bold{x})$, since $\pi_{\text{stable}}$ is guaranteed to be safe starting from $\bold{x}'$, so we can continue to guarantee safety. Otherwise, we use $\pi_{\text{stable}}(\bold{x})$ (which is guaranteed to be safe).

Constructing a stable controller $\pi_{\text{stable}}$ along with stable states $\mathcal{X}_{\text{stable}}$ that is large can be a challenging problem. In our example, the proposed set of stable states (i.e., states where all robots are at rest) has zero volume in the state space. Thus, using shielding with $\pi_{\text{stable}}$ as the backup policy will result in poor performance; indeed, in our example, the robots will never be able to move. The idea behind MPS is to expand the set of states for which we can guarantee safety for an infinite horizon by using a \emph{recovery policy} $\pi_{\text{rec}}$ to try and transition the system to a state $\bold{x}'\in\mathcal{X}_{\text{stable}}$. More precisely, given a state $\bold{x}$, suppose that there exists $T$ such that the finite-horizon trajectory $\zeta=(\bold{x}_0,\bold{x}_1,...,\bold{x}_{T-1})$ generated using $\pi_{\text{rec}}$ from $\bold{x}_0=\bold{x}$ is safe and reaches $\mathcal{X}_{\text{stable}}$ (i.e., $\bold{x}_{T-1}\in\mathcal{X}_{\text{stable}}$). Then, we can guarantee safety starting from $\bold{x}$ by first using $\pi_{\text{rec}}$ for $T-$ steps, and then using $\pi_{\text{stable}}$ afterwards; we call such a state $\bold{x}$ \emph{recoverable}, since it can be recovered to a stable state using $\pi_{\text{rec}}$. Thus, using this combination of $\pi_{\text{rec}}$ and $\pi_{\text{stable}}$ as the backup policy substantially expands the set of states where we can use $\hat{\pi}$.

We can in principle apply this approach to multi-agent systems (with centralized control), where we treat the system as a single high-dimensional system. However, this approach can work poorly---if even a single agent needs to switch to the recovery policy, then every agent must be switched to using the recovery policy. In contrast, our proposed algorithm, multi-agent MPS (MAMPS), considers different choices of policy for different robots. For example, Fig.~\ref{fig:naive_and_ours} shows an instance where the na\"{i}ve approach of treating the system as a single-agent system and using MPS causes all the agents to switch to the recovery policy, whereas our MAMPS algorithm avoids this failure.

\begin{figure*}[t]
\includegraphics[width=\textwidth,height=7cm]{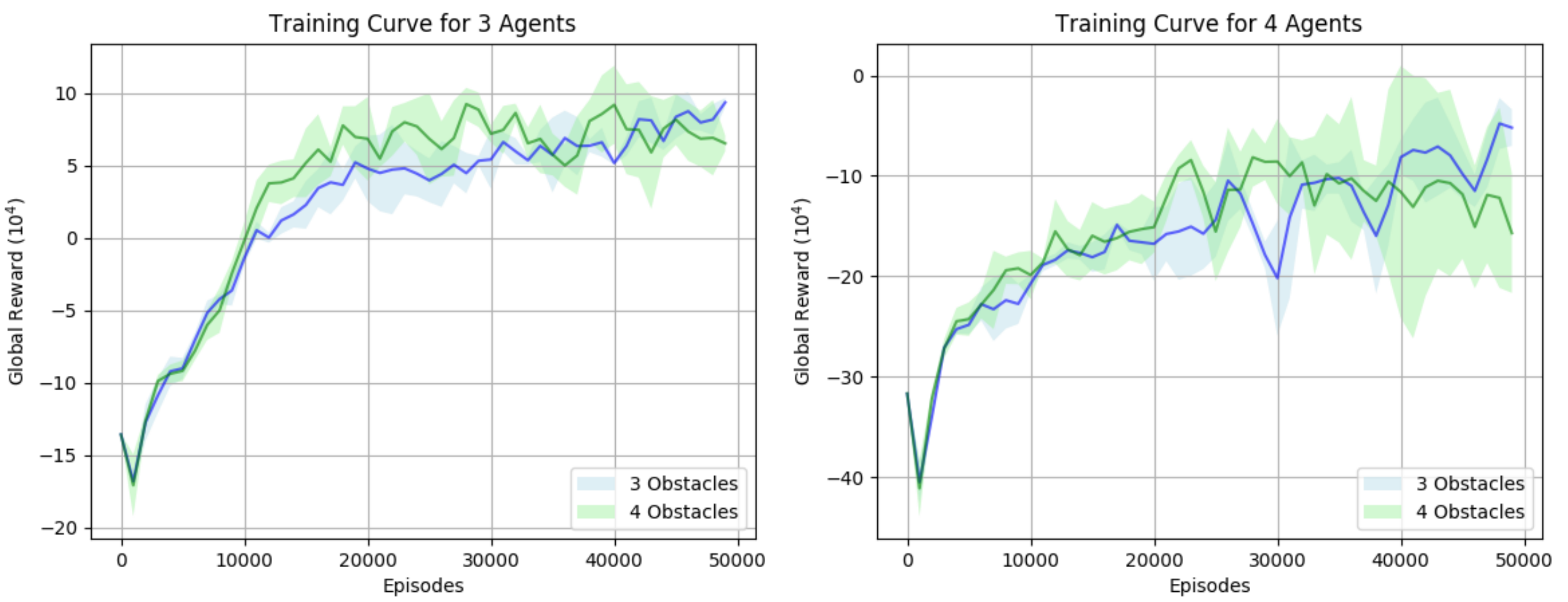}
\caption{\textbf{Training curves for the MADDPG algorithm.} We show the cases of 3 agents (left) and 4 agents (right).}
\label{fig:curve}
\end{figure*}

\textbf{Recovery policy.}
As with MPS, our approach uses a recovery policy $\pi_{\text{rec}}=(\pi_{\text{rec}}^1,...,\pi_{\text{rec}}^N)$ that tries to transition the system from any state $\bold{x}\in\mathcal{X}$ to a stable state $\bold{x}'\in\mathcal{X}_{\text{stable}}$. This policy can be manually specified or learned using reinforcement learning methods. Like $\hat{\pi}$, our algorithm works with any choice of recovery policies. In our example, we can use the policy $\pi_{\text{rec}}^i$ that decelerates the agent as fast as possible until it reaches a stop. Once all robots come to a stop (assuming no robot hits an obstacle or another robot), then the system is in a stable state.

\textbf{Backup policy.}
The backup policy $\pi_{\text{backup}}$ is a combination of the recovery policy and the stable policy. In particular, for agent $i\in[N]$, it uses $\pi_{\text{stable}}^i$ if agent $i$ is stable (i.e., $\bold{x}\in\mathcal{X}_{\text{stable}}^i$), and uses $\pi_{\text{rec}}^i$ otherwise:
\begin{align*}
\pi_{\text{backup}}^i(\bold{x})=\begin{cases}\pi_{\text{stable}}^i(\bold{x})&\text{if}~\bold{x}\in\mathcal{X}_{\text{stable}}^i\\\pi_{\text{rec}}^i(\bold{x})&\text{otherwise},\end{cases}
\end{align*}
for each $i\in[N]$.

\textbf{Recoverable states.}
Specific to the multi-agent setting, we can decompose the notion of recoverability and define recoverability for an individual agent. In particular, a state $\bold{x}\in\mathcal{X}$ is recoverable for agent $i\in[N]$, denoted $\bold{x}\in\mathcal{X}_{\text{rec}}^i$, if there exists $T$ such that the finite-horizon trajectory $\zeta=(\bold{x}_0,\bold{x}_1,...,\bold{x}_{T-1})$ generated from $\bold{x}_0=\bold{x}$ using $\pi_{\text{backup}}$ is (i) safe for agent $i$ (i.e., $\bold{x}_t\in\mathcal{X}_{\text{safe}}^i$ for all $t\in\{0,1,...,T-1\}$), and (ii) reaches a stable state for agent $i$ (i.e., $\bold{x}_{T-1}\in\mathcal{X}_{\text{stable}}^i$). Then, the set of recoverable states is
\begin{align*}
\mathcal{X}_{\text{rec}}=\bigcap_{i=1}^N\mathcal{X}_{\text{rec}}^i.
\end{align*}
We say a state $\bold{x}\in\mathcal{X}_{\text{rec}}$ is \emph{recoverable}. It is easy to see that if $\bold{x}$ is recoverable, then there exists $T\le T_{\text{max}}$ such that the finite-horizon trajectory $\zeta=(\bold{x}_0,\bold{x}_1,...,\bold{x}_{T-1})$ generated from $\bold{x}_0=\bold{x}$ using $\pi_{\text{backup}}$ is (i) safe (i.e., $\bold{x}_t\in\mathcal{X}_{\text{safe}}$ for all $t\in\{0,1,...,T-1\}$), and (ii) reaches a stable state (i.e., $\bold{x}_{T-1}\in\mathcal{X}_{\text{stable}}$).
In this definition, $T_{\text{max}}\in\mathbb{N}$ is a hyperparameter that bounds the length of the trajectory we need to check if recoverability holds, making it feasible to check recoverability in simulation.

In particular, we can check whether a state is recoverable (either for an individual agent or for overall) by simulating $\pi_{\text{backup}}$. This check concludes that $\bold{x}$ is recoverable if and only if $\bold{x}$ is actually recoverable, then $\bold{x}$ is guaranteed to be recoverable. Algorithm~\ref{alg:recovery} performs this check, and returns a vector $\rho=(\rho_1,...,\rho_N)\in\{0,1\}^N$, where $\rho_i\in\{0,1\}$ indicates whether $\bold{x}\in\mathcal{X}_{\text{rec}}^i$ ($\rho_i=1$) or not ($\rho_i=0$).

\textbf{Multi-agent model predictive shielding.}
Our algorithm, multi-agent model predictive shielding (MAMPS),
\footnote{In this name, ``model predictive'' refers to the fact that we are simulating trajectories according to the dynamics model to determine whether to use the learned policy or the backup policy.}
chooses whether to use the learned policy $\hat{\pi}^i$ or the recovery policy $\pi_{\text{rec}}^i$ for each agent $i\in[N]$. In contrast to the MPS approach described above, which either uses $\hat{\pi}^i$ for every agent $i$ or uses $\pi_{\text{rec}}^i$ for every agent $i$, MAMPS considers different choices of learned policy or recovery policy for different agents. We represent the possible choices by a \emph{configuration} $b=(b_1,...,b_N)\in\mathcal{B}=\{0,1\}^N$. In particular, $b_i\in\{0,1\}$ indicates whether to use $\hat{\pi}^i$ ($b_i=1$) or $\pi_{\text{rec}}^i$ ($b_i=0$). For any $b\in\mathcal{B}$, we use $\pi_b$ to denote the corresponding combination of policies for each agent---i.e.,
\begin{align*}
\pi_b(\bold{x})=\begin{cases}\hat{\pi}^i(\bold{x})&\text{if}~b_i=1\\\pi^i_{\text{backup}}(\bold{x})&\text{otherwise}.\end{cases}
\end{align*}
Our key insight is that we can use $\pi_b$ for \emph{any} configuration $b\in\mathcal{B}$ as long as $f^{(\pi_b)}(\bold{x})\in\mathcal{X}_{\text{rec}}$. Since we can check recoverability of any state $\bold{x}$ using simulation, we can simply enumerate over all configurations $b\in\mathcal{B}$ to find one that satisfies this condition. If there are multiple choices of $b$, then we want to choose the one that maximizes $\|b\|_1$---i.e., the one that maximizes the number of agents $\hat{\pi}$ using their learned control policy. Thus, we want to compute
\begin{align}
\label{eqn:optimization}
b^*=\operatorname*{\arg\max}_{b\in\mathcal{B}}\|b\|_1\cdot\mathbb{I}[f^{(\pi_b)}(\bold{x})\in\mathcal{X}_{\text{rec}}],
\end{align}
where $\mathbb{I}$ is the indicator function (taking values in $\{0,1\}$).

For systems with many agents, iterating over all combinations $b\in\mathcal{B}$ can become very expensive, since $|\mathcal{B}|=2^N$ is exponential in the number of agents. Especially for systems with limited computational resources, computing $b^*$ can be intractable. Thus, MAMPS instead solves  (\ref{eqn:optimization}) approximately using a greedy iterative search strategy. In particular, MAMPS starts off by considering the best possible candidate configuration $b=(1,1,...,1)$---i.e., every agent uses the learned policy. Then, it checks whether $f^{(\pi_b)}\in\mathcal{X}_{\text{rec}}$. If so, then we can use $\pi_b$. Otherwise, there are agents for which $\bold{x}$ is irrecoverable---i.e.,
\begin{align*}
\mathcal{I}=\{i\in[N]\mid\bold{x}^i\not\in\mathcal{X}_{\text{rec}}^i\}\neq\varnothing.
\end{align*}
For agents $i\in\mathcal{I}$, MAMPS switches to using the backup policy---i.e., $b_i=0$.

Note that switching an agent from $b_i=1$ to $b_i=0$ may cause a different agent that was previously recoverable to become irrecoverable.
As an example, the illustration of the MAMPS policy in Fig.~\ref{fig:naive_and_ours} shows a case where switching one agent causes another to become irrecoverable. Thus, we have to again check for additional agents that have become irrecoverable. We iteratively perform this process until we find a configuration $b$ such that $f^{(\pi_b)}(\bold{x})\in\mathcal{X}_{\text{rec}}$. Thus, we know that $f^{(\pi_{\text{backup}})}(\bold{x})\in\mathcal{X}_{\text{rec}}$, so we can safely use $\pi_b$.

The full MAMPS algorithm is shown in Algorithm~\ref{alg:shield}. For convenience, when representing $b$, this algorithm uses true instead of $1$, and false instead of $0$. One subtlety is that while $\|b\|_1$ is monotonically decreasing in this algorithm, but it nevertheless get ``stuck'' at some point $b$ that is not guaranteed to be safe, but there are also no agents switching from $b_i=1$ to $b_i=0$. Thus, the algorithm includes a check to see whether $b$ converges. In this case, it sets $b=(0,...,0)$, so $\pi_b=\pi_{\text{backup}}$ is guaranteed to be safe. We have the following guarantee (proved in Appendix~\ref{theorem:3}):

\noindent\textbf{Theorem 1:} The MAMPS policy $\pi_{\text{MAMPS}}$ is safe.

\section{Experiments}
\label{sec:experiments}

In our experiments, we aim to answer the following research questions:
\begin{itemize}
\item How does MAMPS compare to using the learned policy without any shield (in terms of reward and safety)?
\item How does MAMPS compare to the na\"{i}ve shield (i.e., treat the system as a single-agent system and use MPS)?
\item How does the performance of MAMPS vary with respect to the number of agents or obstacles?
\end{itemize}
All experiments are performed on a server with an Intel Xeon Gold 6148 CPU an Nvidia RTX 2080 Ti GPUs.

\begin{figure*}[h]
\includegraphics[width=\textwidth,height=7cm]{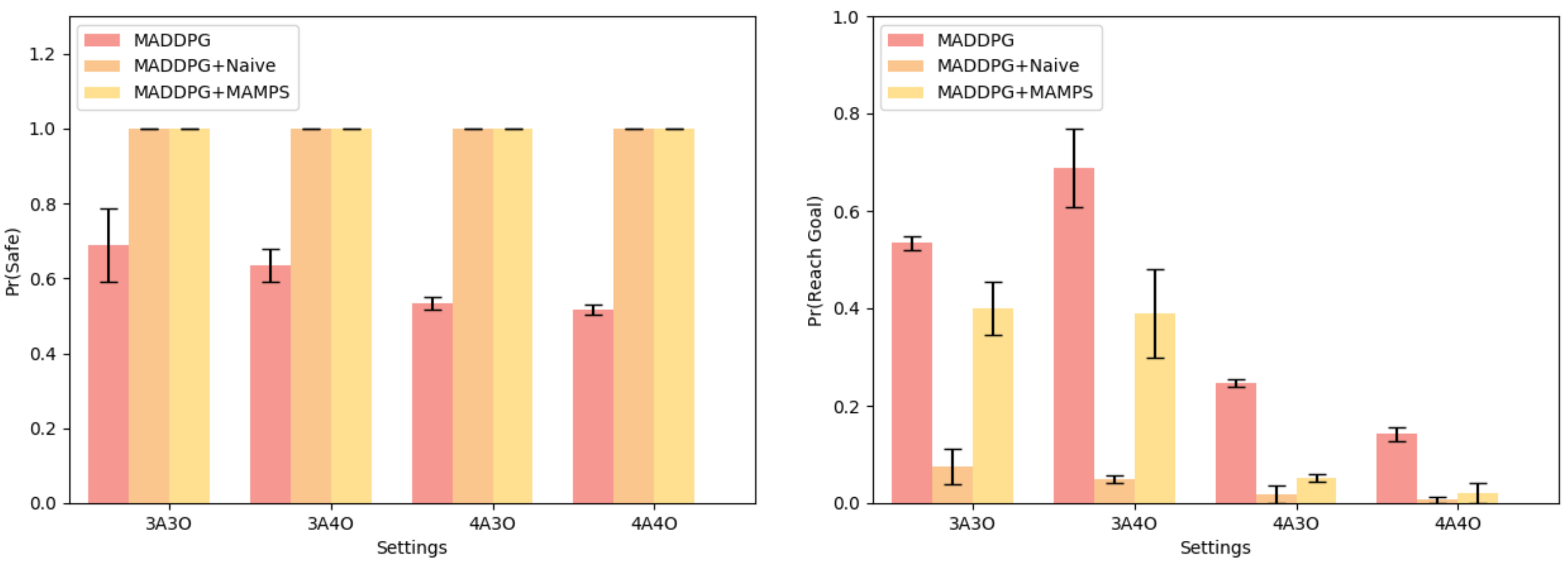}
\caption{\textbf{Probability of safety (left) and probability of reaching the goal (right) for multi-agent systems.} Here, ``3A3O", ``3A4O", "4A3O", and ``4A4O" represent ``3 agents and 3 obstacles", ``3 agents and 4 obstacles", ``4 agents and 3 obstacles", and ``4 agents and 4 obstacles", respectively.}
\label{fig:three_Approach}
\end{figure*}

\subsection{Setup}

We perform our experiments using the multi-agent particle environment~\cite{mordatch2017emergence}. This environment consists of a multi-agent system in which a set of $N$ agents is trying to reach a set of $N$ goals. In our setup, each agent is assigned a specific goal that they are trying to reach. There is also a collection of $M$ obstacles that the agents must avoid; in addition, the agents must avoid colliding with one another.

The learned policy $\hat{\pi}$ is trained using multi-agent deep deterministic policy gradients (MADDPG)~\cite{lowe2017multi}, which uses decentralized actors and a centralized critic. Our reward function is
\begin{align*}
r(\mathbf{x},\mathbf{u}) &= r_{\text{goal}}(\mathbf{x},\mathbf{u}) + \lambda\cdot r_{\text{bonus}}(\mathbf{x},\mathbf{u}) + \lambda'\cdot r_{\text{safe}}(\mathbf{x},\mathbf{u}),
\end{align*}
where the reward for approaching goals is
\begin{align*}
r_{\text{goal}}(\mathbf{x},\mathbf{u})=-\sum_{i=1}^Nd(\bold{x}^i,\bold{g}^i),
\end{align*}
the bonus reward for arriving at a goal is
\footnote{We find that MADDPG does not perform well without this term.}
\begin{align*}
r_{\text{bonus}}(\mathbf{x},\mathbf{u})=\sum_{i=1}^N\mathbb{I}(d(\bold{x}^i, \bold{g}^i) \le \epsilon),
\end{align*}
the penalty for collisions is
\begin{align*}
r_{\text{safe}} =& -\sum_{i=1}^N\sum_{j=i+1}^N \mathbb{I} \left(d(\bold{x}^i, \bold{x}^j) < 2 \cdot r_\text{robot} + m \right) \\
&\hspace{0.2in} - \sum_{i=1}^N\sum_{j=1}^M \mathbb{I} \left(d(\bold{x}^i, \bold{z}^m) < r_\text{robot} + r_\text{obstacle} + m \right),
\end{align*}
$\lambda,\lambda',\epsilon\in\mathbb{R}_{>0}$ are hyperparameters, and $\mathbb{I}(C)\in\{0,1\}$ is the indicator function, which indicates whether condition $C$ holds. Parameter choices are shown in Appendix~\ref{sec:parameters}.  

For the recovery policy $\pi_\text{rec}$, we use the policy that applies the maximum possible acceleration to decrease the robot's velocity (i.e., apply the brakes as hard as possible). Note that if the robot uses recovery policy while it is at rest ($v=0$), then it will stay in the same state. Better choices may be possible, but we find that this simple choice is very effective.


\subsection{Results}

In Fig.~\ref{fig:curve}, we show learning curves for the MADDPG algorithm. As can be seen, it successfully learns how to control each system. 
In Fig.~\ref{fig:three_Approach}, we compare our approach, MADDPG with MAMPS (MADDPG+MAMPS), with two baselines: (i) using just the learned policy (MADDPG), and (ii) using MADDPG with the na\"{i}ve shield (MADDPG+Na\"ive). For completeness, we have show the na\"{i}ve algorithm in Appendix~\ref{sec:naive} Our results are averaged over 500 episodes. We compare two metrics: probability of safety (left) and probability of reaching the goal (right). For safety, we measure the fraction of the agent/episodes pairs for which safety is ensured for the entire duration of the episode---i.e.,
\begin{align*}
\text{Pr}(\text{safe}) \approx \frac{1}{N\cdot K}\sum_{i=1}^K\sum_{j=1}^N\delta_{ij}^{\text{safe}},
\end{align*}
where $K=500$ is the number of episodes, $\delta_{ij}^{\text{safe}}\in\{0,1\}$ indicates whether agent $j$ is safe during the entirety of episode $i$ (i.e., $\bold{x}_t^i\in\mathcal{X}_{\text{safe}}^i$). For reaching the goal, we count the fraction of agent/episode pairs for which the agent reaches the goal at any point during the episode---i.e.,
\begin{align*}
\text{Pr}(\text{reach goal}) \approx \frac{1}{N\cdot K}\sum_{i=1}^K\sum_{j=1}^N\delta_{ij}^{\text{goal}},
\end{align*}
where $\delta_{ij}^{\text{goal}}\in\{0,1\}$ indicates whether agent $j$ reaches its goal at any point in episode $i$.

\textbf{Ensuring safety.}
Note that MADDPG alone performs quite poorly in terms of safety---it can guarantee safety less than 50\% of the time when there are 4 agents and 4 obstacles. The poor performance in terms of safety happens even though MADDPG includes a penalty for collisions. The difficulty is that there are a huge number of possible configurations of the state space, and it is not possible to ensure that MADDPG trains the neural network to account for all of them. As expected, for MADDPG, the probability of safety decreases as there are more agents or more obstacles. Finally, both MADDPG+MAMPS and MADDPG+Na\"ive guarantee the safety of multi-agent for all agents in all episodes.

\textbf{Reaching goals.}
As expected, MADDPG+MAMPS achieves its goals less frequently than MADDPG since MADDPG is allowed to have unsafe collisions without affecting this metric. More interestingly, MADDPG+MAMPS substantially outperforms MADDPG+Na\"ive in terms of performance, often by an order of magnitude. Furthermore, the performance of MADDPG+MAMPS is quite close to the performance of MADDPG alone in settings where there are 3 agents. The relative performance degrades substantially when there are 4 agents, likely because agent-agent collisions increase significantly, which causes agents to use the recovery policy, and therefore fail to reach their goals.

\textbf{Discussion.}
Overall, these results demonstrate the substantial promise of the MAMPS approach. In some settings, it is able to guarantee safety without sacrificing very much performance. There is inevitably some tradeoff between safety and good performance. Nevertheless, we believe that there is much potential to improve MAMPS and reduce how much performance must be sacrificed to ensure safety.

\section{Conclusion}
\label{sec:discussion}

We have proposed a novel algorithm, MAMPS, for ensuring the safety of a learned control policy for multi-agent systems. Our algorithm comes with strong theoretical guarantees on safety. Furthermore, our experimental results show how MAMPS can ensure safety without sacrificing much performance, and that MAMPS can substantially outperform a na\"{i}ve approach. There is much room for future work---e.g., allowing for partially observed environments and closing the gap in performance between MAMPS and the learned policy.



\section*{APPENDIX}

\subsection{Proof of Theorem 1}
\label{theorem:3}

Note that to prove the theorem statement, it suffices to prove that $\bold{x}\in\mathcal{X}_{\text{rec}}$, then $f^{(\pi_{\text{shield}})}(\bold{x})\in\mathcal{X}_{\text{rec}}$ as well. In particular, by induction, this claim implies that a trajectory $(\bold{x}_0,\bold{x}_1,...)$ generated using $\pi_{\text{shield}}$ from initial state $\bold{x}_0\in\mathcal{X}_{\text{rec}}$ satisfies $\bold{x}_t\in\mathcal{X}_{\text{rec}}$ for all $t\ge0$. Furthermore, by definition we have $\mathcal{X}_{\text{stable}}\subseteq\mathcal{X}_{\text{rec}}\subseteq\mathcal{X}_{\text{safe}}$. Thus, a trajectory $(\bold{x}_0,\bold{x}_1,...)$ generated using $\pi_{\text{shield}}$ starting from $\bold{x}_0\in\mathcal{X}_{\text{stable}}$ satisfies $x(t)\in\mathcal{X}_{\text{safe}}$---i.e., $\pi_{\text{shield}}$ is safe.

Next, we prove the remaining claim. Consider the action $\bold{u}=\pi_b(\bold{x})$ returned by Algorithm~\ref{alg:shield}. Since Algorithm~\ref{alg:shield} checks that $\bold{x}'=f^{(\pi_b)}(\bold{x})\in\mathcal{X}_{\text{rec}}$, we are guaranteed that $\bold{x}'$ is recoverable. The challenge is proving that Algorithm~\ref{alg:shield} actually returns an action.

First, we show that for the choice $b=(0,...,0)$, we have $f^{(\pi_b)}(\bold{x})\in\mathcal{X}_{\text{rec}}$. To this end, note that in this case, we have $\pi_b=\pi_{\text{backup}}$. Then, note that since $\bold{x}$ is recoverable, we know that there exists $T\in\{0,1,...,T_{\text{max}}-1\}$ such that the trajectory $(\bold{x}_0,\bold{x}_1,...,\bold{x}_T)$ generated using $\pi_{\text{backup}}$ from $\bold{x}_0=\bold{x}$ satisfies the following: (i) it is safe (i.e., $\bold{x}_t\in\mathcal{X}_{\text{safe}}$ for all $t\in\{0,1,...,T\}$), and (ii) it reaches $\mathcal{X}_{\text{stable}}$ (i.e., $\bold{x}_T\in\mathcal{X}_{\text{stable}}$). Thus, for $\bold{x}'=f^{(\pi_{\text{backup}})}(\bold{x})$, consider the trajectory $(\bold{x}_1,...,\bold{x}_T)$ generated using $\pi_{\text{backup}}$ from $\bold{x}_1=\bold{x}'$. Trivially, this trajectory is also safe and reaches $\mathcal{X}_{\text{stable}}$. Thus, $\bold{x}'$ is recoverable, as claimed. Finally, the check in Algorithm~\ref{alg:shield} eventually considers $b=(0,...,0)$; thus, it is guaranteed to terminate. The claim follows. $\blacksquare$

\subsection{Parameters}
\label{sec:parameters}

\textbf{Environment parameters.}
Maximum episode length: 300. Total size of the environment: $[-1,1]^2$. Object dimensions: $r_\text{robot}=0.1$, $r_\text{obstacle}=0.1$, $r_\text{goal}=0.05$. Allowed accelerations: $[-1,1]$. Allowed velocities: $[0,3]$. Time step: $0.025$.

\textbf{MADDPG parameters.}
Learning rate: $10^{-3}$. Discount factor $\gamma$: 0.95. Size of minibatch sample: 1024. Actor and critic networks: 8 fully connected layers with 128 hidden units each for three-agent settings and 10 fully connected layers with 128 hidden units each for four-agent settings.

\textbf{MAMPS parameters.}
Maximum trajectory length $T_{\text{max}}$ for recovery check: $120$.

\subsection{Na\"{i}ve Approach}
\label{sec:naive}

Algorithm~\ref{alg:naive_shield} shows the na\"{i}ve approach of treating the system as a single-agent system, and then using MPS.

\begin{algorithm}[t]
\caption{Control policy using the na\"{i}ve approach.}
\label{alg:naive_shield}
 \begin{algorithmic}[1]
 \renewcommand{\algorithmicrequire}{\textbf{function}}
 \REQUIRE 
  Na\"iveShield($\bold{x}$):
  \IF{IsRecoverable($f^{(\hat{\pi})}(\bold{x})$)}
  \RETURN $\hat{\pi}(\bold{x})$
  \ELSE
  \RETURN $\pi_\text{rec}(\bold{x})$
  \ENDIF
\renewcommand{\algorithmicensure}{\textbf{Output:}} 
\end{algorithmic} 
\end{algorithm}

\bibliographystyle{IEEEtran.bst}

\bibliography{ref.bib}

\begin{thebibliography}{10}
\providecommand{\url}[1]{#1}
\csname url@rmstyle\endcsname
\providecommand{\newblock}{\relax}
\providecommand{\bibinfo}[2]{#2}
\providecommand\BIBentrySTDinterwordspacing{\spaceskip=0pt\relax}
\providecommand\BIBentryALTinterwordstretchfactor{4}
\providecommand\BIBentryALTinterwordspacing{\spaceskip=\fontdimen2\font plus
\BIBentryALTinterwordstretchfactor\fontdimen3\font minus
  \fontdimen4\font\relax}
\providecommand\BIBforeignlanguage[2]{{%
\expandafter\ifx\csname l@#1\endcsname\relax
\typeout{** WARNING: IEEEtran.bst: No hyphenation pattern has been}%
\typeout{** loaded for the language `#1'. Using the pattern for}%
\typeout{** the default language instead.}%
\else
\language=\csname l@#1\endcsname
\fi
#2}}

\bibitem{gu2017deep}
S.~Gu, E.~Holly, T.~Lillicrap, and S.~Levine, ``Deep reinforcement learning for
  robotic manipulation with asynchronous off-policy updates,'' in \emph{2017
  IEEE international conference on robotics and automation (ICRA)}.\hskip 1em
  plus 0.5em minus 0.4em\relax IEEE, 2017, pp. 3389--3396.

\bibitem{andrychowicz2018learning}
M.~Andrychowicz, B.~Baker, M.~Chociej, R.~Jozefowicz, B.~McGrew, J.~Pachocki,
  A.~Petron, M.~Plappert, G.~Powell, A.~Ray, \emph{et~al.}, ``Learning
  dexterous in-hand manipulation,'' \emph{arXiv preprint arXiv:1808.00177},
  2018.

\bibitem{tanner2003coordination}
H.~Tanner, A.~Jadbabaie, and G.~J. Pappas, ``Coordination of multiple
  autonomous vehicles,'' in \emph{IEEE Mediterranean Conference on Control and
  Automation}, 2003, pp. 869--876.

\bibitem{srinivasa2010herb}
S.~S. Srinivasa, D.~Ferguson, C.~J. Helfrich, D.~Berenson, A.~Collet,
  R.~Diankov, G.~Gallagher, G.~Hollinger, J.~Kuffner, and M.~V. Weghe, ``Herb:
  a home exploring robotic butler,'' \emph{Autonomous Robots}, vol.~28, no.~1,
  p.~5, 2010.

\bibitem{levine2013guided}
S.~Levine and V.~Koltun, ``Guided policy search,'' in \emph{International
  Conference on Machine Learning}, 2013, pp. 1--9.

\bibitem{gillula2012guaranteed}
J.~H. Gillula and C.~J. Tomlin, ``Guaranteed safe online learning via
  reachability: tracking a ground target using a quadrotor,'' in \emph{2012
  IEEE International Conference on Robotics and Automation}.\hskip 1em plus
  0.5em minus 0.4em\relax IEEE, 2012, pp. 2723--2730.

\bibitem{akametalu2014reachability}
A.~K. Akametalu, J.~F. Fisac, J.~H. Gillula, S.~Kaynama, M.~N. Zeilinger, and
  C.~J. Tomlin, ``Reachability-based safe learning with gaussian processes,''
  in \emph{53rd IEEE Conference on Decision and Control}.\hskip 1em plus 0.5em
  minus 0.4em\relax IEEE, 2014, pp. 1424--1431.

\bibitem{saha2014automated}
I.~Saha, R.~Ramaithitima, V.~Kumar, G.~J. Pappas, and S.~A. Seshia, ``Automated
  composition of motion primitives for multi-robot systems from safe ltl
  specifications,'' in \emph{2014 IEEE/RSJ International Conference on
  Intelligent Robots and Systems}.\hskip 1em plus 0.5em minus 0.4em\relax IEEE,
  2014, pp. 1525--1532.

\bibitem{fisac2019bridging}
J.~F. Fisac, N.~F. Lugovoy, V.~Rubies-Royo, S.~Ghosh, and C.~J. Tomlin,
  ``Bridging hamilton-jacobi safety analysis and reinforcement learning,'' in
  \emph{2019 International Conference on Robotics and Automation (ICRA)}.\hskip
  1em plus 0.5em minus 0.4em\relax IEEE, 2019, pp. 8550--8556.

\bibitem{liu2017planning}
S.~Liu, M.~Watterson, K.~Mohta, K.~Sun, S.~Bhattacharya, C.~J. Taylor, and
  V.~Kumar, ``Planning dynamically feasible trajectories for quadrotors using
  safe flight corridors in 3-d complex environments,'' \emph{IEEE Robotics and
  Automation Letters}, vol.~2, no.~3, pp. 1688--1695, 2017.

\bibitem{schwarting2017parallel}
W.~Schwarting, J.~Alonso-Mora, L.~Pauli, S.~Karaman, and D.~Rus, ``Parallel
  autonomy in automated vehicles: Safe motion generation with minimal
  intervention,'' in \emph{2017 IEEE International Conference on Robotics and
  Automation (ICRA)}.\hskip 1em plus 0.5em minus 0.4em\relax IEEE, 2017, pp.
  1928--1935.

\bibitem{alshiekh2018safe}
M.~Alshiekh, R.~Bloem, R.~Ehlers, B.~K{\"o}nighofer, S.~Niekum, and U.~Topcu,
  ``Safe reinforcement learning via shielding,'' in \emph{Thirty-Second AAAI
  Conference on Artificial Intelligence}, 2018.

\bibitem{aisafety}
D.~Amodei, C.~Olah, J.~Steinhardt, P.~Christiano, J.~Schulman, and D.~Mané,
  ``Concrete problems in ai safety,'' \emph{arXiv preprint arXiv:1606.06565},
  2016.

\bibitem{berkenkamp2017safe}
F.~Berkenkamp, M.~Turchetta, A.~Schoellig, and A.~Krause, ``Safe model-based
  reinforcement learning with stability guarantees,'' in \emph{Advances in
  neural information processing systems}, 2017, pp. 908--918.

\bibitem{bastani2018verifiable}
O.~Bastani, Y.~Pu, and A.~Solar-Lezama, ``Verifiable reinforcement learning via
  policy extraction,'' in \emph{Advances in Neural Information Processing
  Systems}, 2018, pp. 2494--2504.

\bibitem{ivanov2019verisig}
R.~Ivanov, J.~Weimer, R.~Alur, G.~J. Pappas, and I.~Lee, ``Verisig: verifying
  safety properties of hybrid systems with neural network controllers,'' in
  \emph{Proceedings of the 22nd ACM International Conference on Hybrid Systems:
  Computation and Control}.\hskip 1em plus 0.5em minus 0.4em\relax ACM, 2019,
  pp. 169--178.

\bibitem{DBLP:journals/corr/abs-1803-08552}
\BIBentryALTinterwordspacing
K.~P. Wabersich and M.~N. Zeilinger, ``Linear model predictive safety
  certification for learning-based control,'' \emph{CoRR}, vol. abs/1803.08552,
  2018. [Online]. Available: \url{http://arxiv.org/abs/1803.08552}
\BIBentrySTDinterwordspacing

\bibitem{bastani2019safe}
O.~Bastani, ``Safe reinforcement learning via online shielding,'' \emph{arXiv
  preprint arXiv:1905.10691}, 2019.

\bibitem{li2019robust}
\BIBentryALTinterwordspacing
S.~Li and O.~Bastani, ``Robust model predictive shielding for safe
  reinforcement learning with stochastic dynamics.'' [Online]. Available:
  \url{https://obastani.github.io/docs/rmps.pdf}
\BIBentrySTDinterwordspacing

\bibitem{desai2001modeling}
J.~P. Desai, J.~P. Ostrowski, and V.~Kumar, ``Modeling and control of
  formations of nonholonomic mobile robots,'' 2001.

\bibitem{alonso2015multi}
J.~Alonso-Mora, S.~Baker, and D.~Rus, ``Multi-robot navigation in formation via
  sequential convex programming,'' in \emph{2015 IEEE/RSJ International
  Conference on Intelligent Robots and Systems (IROS)}.\hskip 1em plus 0.5em
  minus 0.4em\relax IEEE, 2015, pp. 4634--4641.

\bibitem{AlonsoMora2010OptimalRC}
J.~Alonso-Mora, A.~Breitenmoser, M.~Rufli, P.~A. Beardsley, and R.~Siegwart,
  ``Optimal reciprocal collision avoidance for multiple non-holonomic robots,''
  in \emph{DARS}, 2010.

\bibitem{rvo}
J.~van~den Berg, M.~Lin, and D.~Manocha, ``Reciprocal velocity obstacles for
  real-time multi-agent navigation,'' 05 2008, pp. 1928--1935.

\bibitem{khan2019learning}
A.~Khan, C.~Zhang, S.~Li, J.~Wu, B.~Schlotfeldt, S.~Y. Tang, A.~Ribeiro,
  O.~Bastani, and V.~Kumar, ``Learning safe unlabeled multi-robot planning with
  motion constraints,'' \emph{arXiv preprint arXiv:1907.05300}, 2019.

\bibitem{lowe2017multi}
R.~Lowe, Y.~Wu, A.~Tamar, J.~Harb, P.~Abbeel, and I.~Mordatch, ``Multi-agent
  actor-critic for mixed cooperative-competitive environments,'' \emph{Neural
  Information Processing Systems (NIPS)}, 2017.

\bibitem{mordatch2017emergence}
I.~Mordatch and P.~Abbeel, ``Emergence of grounded compositional language in
  multi-agent populations,'' \emph{arXiv preprint arXiv:1703.04908}, 2017.

\end{thebibliography}

\end{document}